# MACHINERY OF FUNCTIONAL DYNAMICS IN NATIVE PROTEINS

**Canan Baysal[†] and Ali Rana Atilgan[‡]**

[†]*Laboratory of Computational Biology, Faculty of Engineering and Natural Sciences, Sabanci University, Tuzla 34956, Istanbul, Turkey.*

[‡]*School of Engineering, Bogazici University, Bebek 34342, Istanbul, Turkey.*

We provide evidence that the energy landscapes of folded proteins do not shift with temperature, but the onset of functional dynamics is associated with its effective sampling. The motion of the backbone is described by three distinct regimes. One is associated with slow time scales of the activity along the envelope of the energy surface defining the folded protein. Another, with fast time scales, is due to activity along the pockets decorating the folded-state envelope. The intermediate regime emerges at temperatures where jumps between the pockets become possible, leading to an active protein.

Understanding the temperature dependence of the internal motions in folded proteins is essential for controllability and designability. Recently, a substantial amount of experimental[1-5] and theoretical[6-10] research effort have been invested in this direction. It has now been established that the protein and a solvent shell of thickness ca. 4 Å[11] constitute the system of interest.[9] An essential part of the overall motion of this system is slaved to the bulk solvent, implying that the enthalpic and entropic contributions to the free energy fluctuations in the protein originate in the solvent and the protein-hydration shell system, respectively.[5] Investigations on the temperature dependent properties of folded proteins reveal that hydrated proteins show a dynamical transition above which temperature they are active.[2] Experimentally, the transition temperature is quantified by, for example, monitoring the average fluctuations of hydrogen atoms in neutron scattering. Typical temperatures studied range from well below the transition temperature (which is observed at ca. 190-220 K for different proteins), to temperatures right below unfolding. The fluctuations are anharmonic even at temperatures as low as 100 K.[12] In dehydrated powder samples or for samples in solvents that do not permit functionality, the fluctuation amplitude increases linearly throughout the temperature window studied. The slope of these curves are identified as the force constant of the elastic body studied.[13] In other samples, the slope shows an increase during the transition temperature with an increase in the force constant at physiological temperatures. It has been argued that the dynamical transition is controlled by the solvent.[3,14,15] However, it is far from clear how the solvent effects the landscape, and which processes are onset with increasing temperature.

To gain insight into the processes involved, we study the equilibrium fluctuations and the relaxation phenomena of the fluctuation vector attached to the $C_\alpha$ atoms of the bovine pancreatic trypsin inhibitor (BPTI) by molecular dynamics (MD). We confine our attention to the protein-water system. We report results from simulations on BPTI, using the Protein Data Bank[16] code 5pti[17] as the initial structure with a 6 Å thick hydration water layer at 13 different temperatures in the range 150 – 320 K. The details of the simulations are given elsewhere.[8] We record trajectories of length 2.0 – 4.8 ns depending on the temperature; as the size of the fluctuations increase at higher temperatures, more data are needed for a more reliable analysis. Each 400 ps portion of the trajectories is treated as a separate sample in the analysis. We thus have 5 – 12 data sets at each temperature, and the calculated quantities are averaged over these.

The fluctuation vector for the *i*th residue, $\Delta \mathbf{R}_i$, is calculated as the difference of the actual atomic coordinates from those of a suitably determined average structure calculated for each sample data set, such that only the contributions from the motions of the internal coordinates are included.[8] Our analysis of the data reveals that around the folded state, residue fluctuations, $<\Delta \mathbf{R}_i \cdot \Delta \mathbf{R}_i>$, display the same characteristics irrespective of temperature. Examples are shown for the $C_\alpha$ atoms at temperatures below the protein dynamical transition temperature (149 K), during the transition (209 K), above the transition temperature (265 K), and right below the unfolding temperature (310 K); fluctuations from NMR experiments conducted at 310 K, averaged over the 20 best structures, are also presented for comparison.

We find that the details of the residue-by-residue fluctuations are very similar, although the magnitude of the fluctuations differ markedly. The common feature of the systems at all these temperatures is the average structure of the protein about which the fluctuations occur, an equilibrium property. This is a coarse approach to the analysis of the wealth of data produced by MD and will not give sufficient information on the details of relaxation behavior, a dynamical phenomenon. We characterize the motion of the fluctuation vector by a relaxation function which we term dynamic flexibility[8]

$$C(t) = \frac{\overline{\langle \Delta \mathbf{R}(0) \cdot \Delta \mathbf{R}(t) \rangle}}{\overline{\langle \Delta R^2 \rangle}} \qquad (1)$$

where the bar denotes the average over all residues and the brackets represent the time average over all time windows of length $t$ in the trajectory. In the present study, eq. 1 gives the details of the relaxation phenomena in the part of the potential energy surface restricted to the vicinity of the folded structure. The relatively large scale nature of these motions has contributions from different processes such as harmonic motions between bonded atoms, solvent effects, transitions between conformational substates of the backbone and side chains, as well as larger cooperative fluctuations occurring in, for example, the loop regions.

The overall relaxation may be thought of as the superposition of many different homogeneous processes with different relaxation times;† their collective effect on relaxation will be observed as heterogeneous dynamics,[19]

---

† Note that an exponential decay of all contributing processes are assumed here. In the sub-picosecond regime, non-exponential contributions to relaxation phenomena, probably encompassing quantum mechanical effects, are known to exist.[18] Such time scales and motions are beyond the capacity of the current MD simulations, and are assumed to be averaged out at the time and length scales explored by the $C_\alpha$ fluctuation dynamics.

$$C(t) = \sum_{i=1}^{n} \exp\left(\frac{-t}{\tau_i}\right) \qquad (2)$$

which may be approximated by the stretched exponential, or the Kohlrausch-Williams-Watts (KWW) function[20,21]

$$C(t) = \exp\left(\frac{-t}{\tau_e}\right)^{\beta} \qquad (3)$$

Here $\tau_e$ is a temperature dependent effective relaxation time representing the average contribution of all the processes affecting the relaxation of the fluctuation vector. $\beta$ ($0 \leq \beta \leq 1$) reflects the complexity of the processes involved. It is 1 for a single process characterized by a simple exponential decay and it tends to decrease from 1, not only if a larger number of contributing processes ($n > 1$) are at play, but also if these processes have time and/or length scales spanning different orders of magnitudes.

Whereas it is evident $\tau_e$ in eq. 3 is a dynamical parameter, more subtle is the association of $\beta$ as a thermodynamic quantity.[8] Fluctuations in the thermodynamical entities such as the energy or volume can be used to depict phase transitions. Their characterization can be achieved by monitoring the relevant susceptibilities such as the heat capacity or isothermal compressibility. Since $\beta$ was used to describe the protein dynamical transition,[8] we compare its temperature dependence to that of the system heat capacity in figure 2. Both data may be represented by two plateau regions with a transition between them in the same temperature range. The best-fitting curves are drawn through the $\beta$ data using a Boltzmann sigmoidal function given by

$$\beta = \beta_o + \frac{\beta_1 - \beta_o}{1 + \exp\left(\frac{T_o - T}{\lambda}\right)} \qquad (4)$$

Here, $\beta_o$ and $\beta_1$ are the values before and after the transition, $\lambda$ is a constant governing the slope of the rise during the transition; a similar expression can be written for the heat capacity data. $\lambda$ is found to be 8.0 and 7.4, and the transition temperature $T_o$ is predicted from the inflection point as 193±5 K and 197±4 K from the heat capacity and stretch exponent data, respectively. Hence, the two curves can be said to describe the same type of phenomena, and that we can regard $\beta$ as a type of susceptibility. Furthermore, the heat capacity reflects the added contributions of the *n* individual processes (eq. 2) to the overall energy fluctuations.

These arguments demonstrate that the dynamical aspects of residue fluctuations involve distinct events below and above the transition, with a gradual (over a temperature range of 50 K) onset of the new events as temperature increases. Our analysis of one variable (residue fluctuations) in two separate ways so far leads to two seemingly contradictory results. Equilibrium fluctuations, $<\Delta R^2>$, have the same residue-by-residue details (figure 1), leading one to assume that the same processes govern at all temperatures, despite the presence of a dynamical transition (the latter is also depicted in the inset to figure 2b by the average $C_\alpha$ fluctuations as a function of temperature). Relaxation phenomena, on the other hand, point to the presence of separate processes with different relaxation profiles below and above the dynamical transition (figure 2).

How is it possible that the dynamical transition that leads to a nonlinear increase in the size of the fluctuations and a functional protein at high temperatures is not manifested in the details of the residue-by-residue fluctuations? And how is it that increasing the temperature and therefore introducing more processes into the dynamics leads to "less complicated" dynamics, implied by the increased value of $\beta$ (0.4 above the transition as opposed to 0.2 below)?

A simple way to obtain complex relaxation described by the KWW form of eq. 3 is to include two processes in eq. 2 with somewhat separate time scales, e.g. two orders of magnitude, 1 and 100 ps (too much separated time scales will lead to the averaging out of the faster one upon superposition of the two events). Let us also consider a third event with an intermediate time scale (e.g. 10 ps). The superimposed relaxation curves of the two former events, and all three events together, obtained from eq. 2, are shown in figure 3. Also displayed are the curve fits to intermediate times using eq. 3. We find that the introduction of the transitional event increases $\beta$ from ca. 0.2 to ca. 0.4, while decreasing $\tau_e$; i.e. it leads to a faster relaxation, somewhat closer to a simple exponential decay.

Using the landscape perspective of proteins, we associate the processes governed by the slow time scale (~100 ps) with the motion of the overall protein along the envelope of the energy surface defining the folded structure. The processes with the fast time scale (~1 ps) correspond to the vibrations of the atoms in space, whose more collective manifestations may be monitored through the librational motions of the torsional angles. Their exact form is determined by the pockets decorating the folded-state envelope. The jumps between the pockets are made possible when the molecule gains enough energy, through the heat bath provided by the surrounding medium, to overcome the small energy barriers between them. The frequency of these jumps controls the intermediate regime which becomes accessible once the barriers can be surmounted. The related time scales are intermediate to the other two regimes (~10 ps).

The intermediate regime must correspond to one where local collectiveness operates in such a way that the protein becomes functional without changing the overall energy landscape of the molecule (due to figure 1). Conformational jumps in the torsional angles in the side chains of the surface residues possess both properties; jumps in the backbone torsions, on the other hand, are not good candidates as changes in their rotameric states require the cooperative rearrangement of adjacent dihedral angles and may lead to a shift in the energy landscape. In figure 4 we display representative torsional angle trajectories of Lys15 which is in the binding loop of BPTI in its complex with β-trypsin. The local fluctuations in the backbone, exemplified by the $\psi$ angle, increase with temperature without a change in the rotameric state. Note that at 310 K, right below the unfolding temperature, attempted rotameric jumps on the backbone are hindered by the landscape, as exemplified by the short-lived jump around 1050 ps. Such changes in the rotameric states of the backbone become accessible with unfolding (data at 320 K not shown). Side chain torsional angles, on the other hand, display rotameric jumps at temperatures corresponding to the onset of the dynamical transition, e.g. at 209 K, as shown by the side chain $\chi_3$ angle trajectories in figure 4. Note that the librations and jumps are coupled processes, and the former facilitate the faster relaxation of the isomeric jumps by localizing the spatial reorientation of chain segments.[22]

The analysis outlined above provides a simplistic view of the dynamics of a folded protein below temperatures that will induce unfolding. The forces holding together the native protein and its hydration shell are strong enough such that every detail of the same average structure is maintained irregardless of the input of energy from the environment. Thus, the energy landscape does not change with temperature. The onset of short-range cooperative motions at the dynamical transition is associated with the insertion of an intermediate time scale.

We hypothesize that the same behavior will be observed in all solvents that maintain a similar hydrogen bonding pattern to water.[23] The transition temperature, however, will be shifted depending on the identity of the bulk solvent.[15] Similarly, dehydrated proteins do not show the dynamical transition, because they do not maintain the right conformations on the protein surface, leading to shifts in the energy landscape. More studies in different solvents and on different proteins should be conducted to clarify these points.

The protein is a machine that takes a supply of disordered energy from the environment and converts it into an organized set of molecular motions such that the loop regions will sample a restricted set of conformations, including those essential for biological activity. The motility gained along the proper landscape, guaranteed by the protein-solvent system, with the activation of the intermediate regime at relatively elevated temperatures, is necessary for the protein to become functional. Once the function (e.g. binding) is realized however, there are shifts in the energy landscape that affect the stability of the protein[24] so as to relieve the strain on the system through the shortest possible path.[25]

Partial support from Sabanci University Internal Research Grant A0003-00171 and Bogazici University Research Foundation Project 02R102 are gratefully acknowledged.

**FIGURE CAPTIONS**

*Figure 1*. Fluctuations of the $C_\alpha$ atoms, $<\Delta \mathbf{R}_i \cdot \Delta \mathbf{R}_i>$, (a) well below the dynamic transition temperature (149 K); (b) during the transition (209 K); (c) above the transition temperature (265 K); (d) at the temperature where the NMR experiments are conducted (310 K); (e) averaged over the 20 NMR structures reported in the PDB structure 1pit;[26] standard error on the mean is shown as bars. Overall, the residue-by-residue fluctuation patterns are the same at all temperatures and the NMR structures, but note the difference in the absolute value of the fluctuations.

*Figure 2*. (a) System heat capacity, and (b) the stretch exponent β, from the fit of eq. 4 to the relaxation of $C_\alpha$ fluctuations, as a function of temperature. The curves are best-fit of Boltzmann sigmoidal functions to the data points (eq. 4). The goodness of fit, determined by the $R^2$ values are 0.92 in (a) and 0.94 in (b). $\lambda$ is found to be 8.0 and 7.4 and the transition temperature is predicted from the inflection point to be 193±5 K and 197±4 K from the heat capacity and *β* data, respectively. Thermal fluctuations averaged over all residues are also displayed in the inset to (b).

*Figure 3*. Effect of intermediate time scales on the overall dynamics. Assuming there are three processes, each showing single exponential decay, with relaxation times of 1, 10, and 100 ps, the cumulative relaxation behavior of all three processes and the slowest and fastest processes are shown with best-fitted stretch exponential curve fits. *β* = 0.4 and $\tau_e$ = 15 ps for the former, whereas *β* = 0.2 and $\tau_e$ = 35 ps for the latter.

*Figure 4*. Sample torsional angle trajectories of 2 ns duration for Lys15 at the same temperatures as shown in fig. 2. This residue is in the binding loop of BPTI in its complex with β-trypsin. (a) Along the backbone, exemplified here by the $\psi$ angle, the torsions are constrained so that no conformational jumps between rotational isomeric states are observed; the size of the fluctuations, on the other hand, increase above the transition temperature. For example, the average angles and the standard deviations are 78±10° at 149 and 209 K; these become 84±14° and 92±20° at 265 and 310 K, respectively. (b) Along the side-chains, exemplified here by the $\chi_3$ angle, the torsions are more free to move. In fact, conformational transitions are observed even during the dynamical transition. Well below the transition temperature (149 K) the average angles and the fluctuations are 177±7°; at higher temperatures, the main conformer is still the *trans* form with standard deviations of 11, 11, and 15° at 209, 265, and 310 K, respectively.

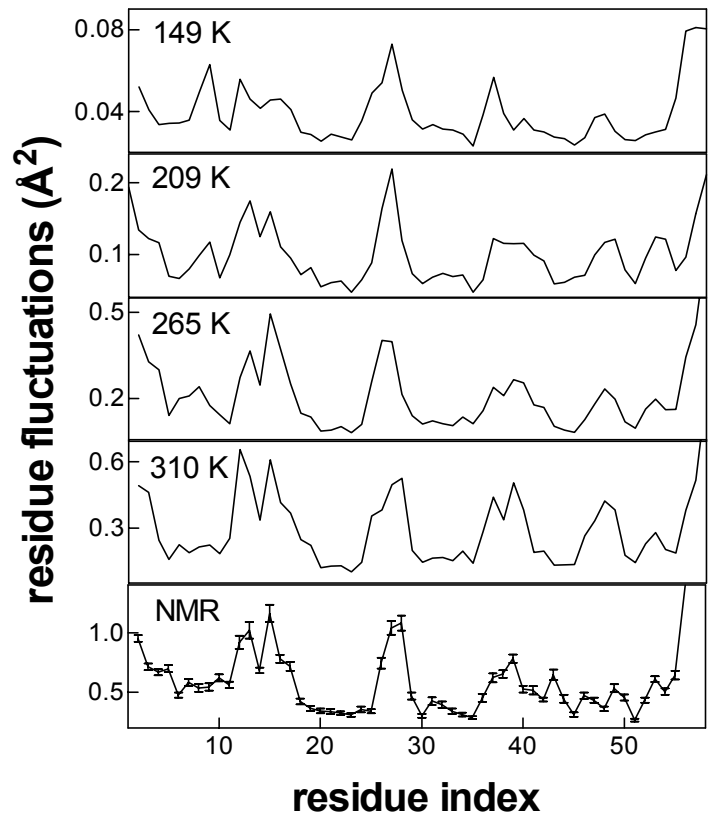

Figure 1

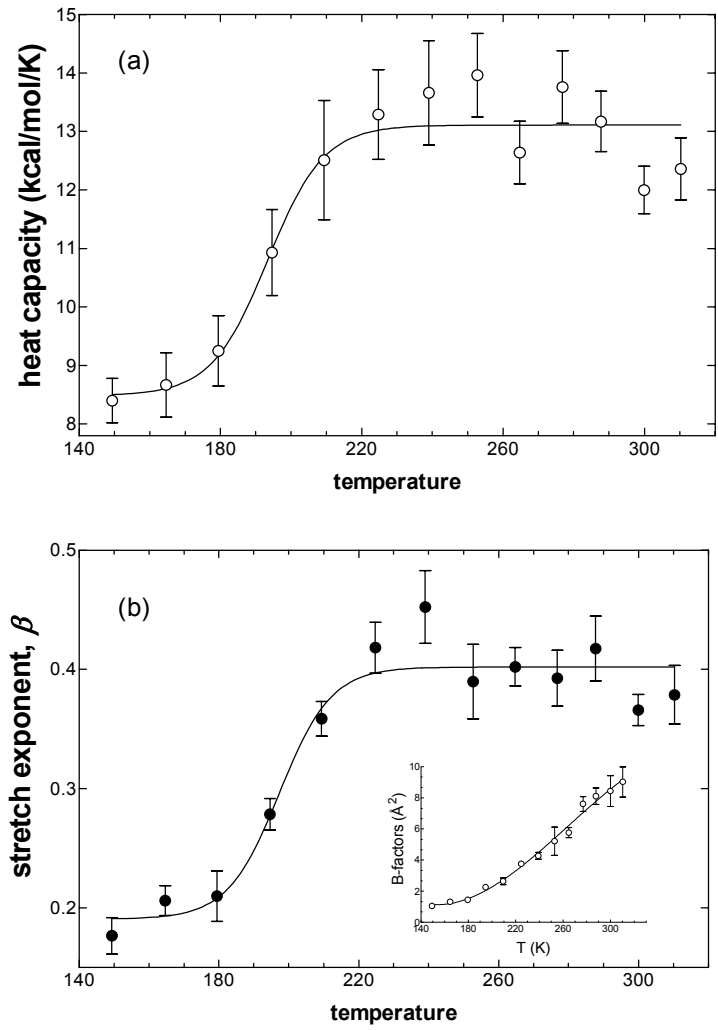

Figure 2

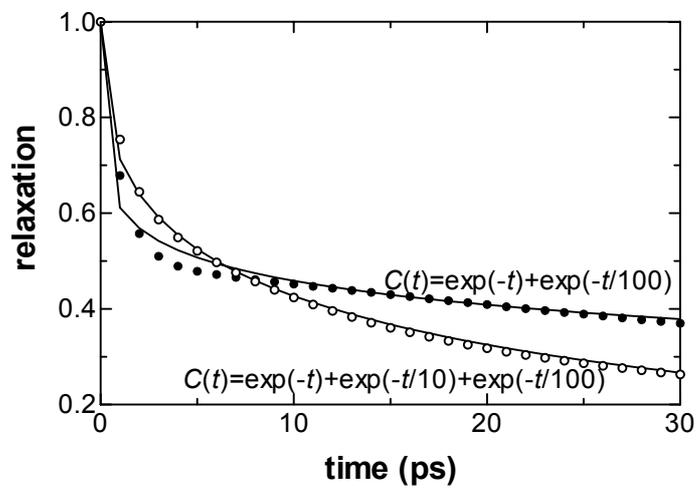

Figure 3

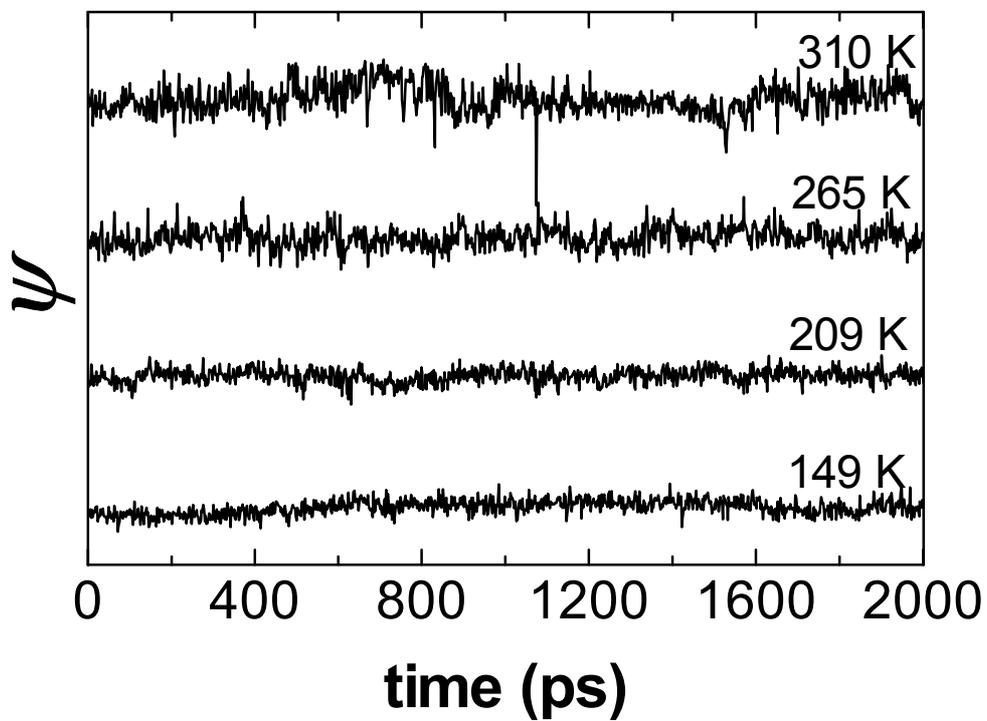
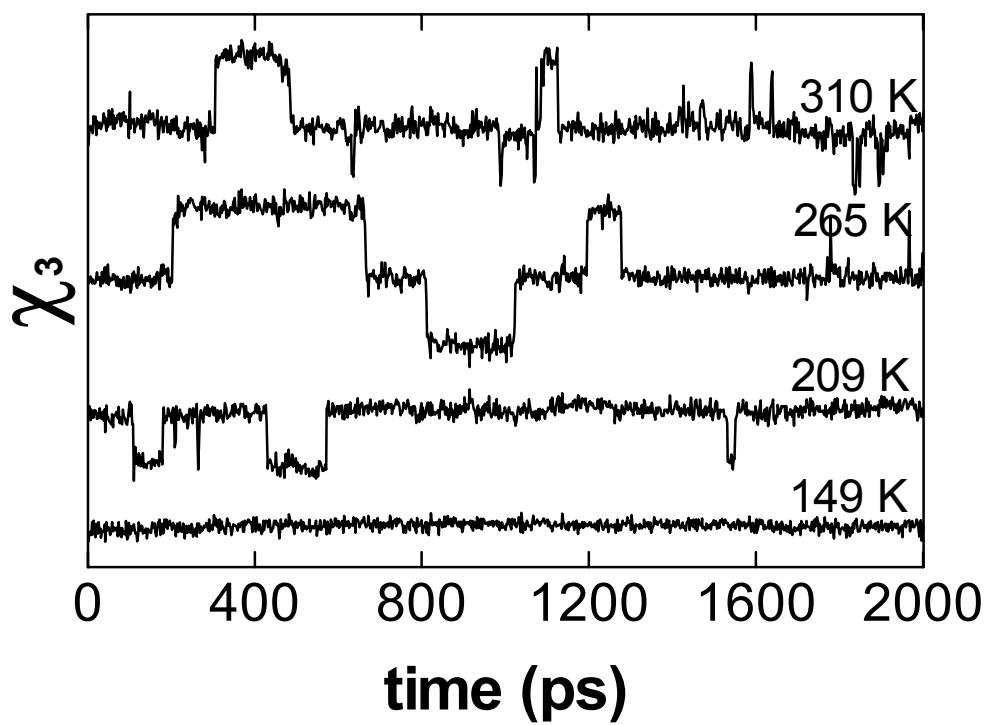

Figure 4

13